# The experimental challenge of detecting solar axion-like particles to test the cosmological ALP-photon oscillation hypotheses

F.T. Avignone III[a,*], R.J. Creswick[a], and S. Nussinov[a,b]

[a] Department of Physics and Astronomy, University of South Carolina, Columbia, SC 29208, USA
[b] Department of Physics, Tel Aviv University, Tel Aviv, Israel

ABSTRACT

We consider possible experimental tests of recent hypotheses suggesting that TeV photons survive the pair production interaction with extragalactic background light over cosmological distances by converting to axion-like particles (ALPs) in galactic magnetic fields. We show that proposed giant ultra-low background scintillation detectors will even have a difficult time reaching the present CAST sensitivity, which is one to two orders of magnitude less sensitive than necessary for a meaningful test of the ALP-photon oscillation hypothesis. Potential alternative tests are briefly discussed.

**INTRODUCTION**

In 2007, De Angelis, Roncadelli and Mansutti (DARM) [1], proposed that the observation of ultra-high energy gamma rays with IACTs (Imaging Atmospheric Cerenkov Telescopes) constituted evidence for a new light spin-zero boson. The basic idea stems from the fact that ultra-high energy photons have a very large cross section, $\sigma(\gamma\gamma \to e^+e^-)$, for pair production via the interaction with the extragalactic background radiation. Accordingly, their observed flux should be significantly less than observed by the H.E.S.S. [2], MAGIC [3], CANGAROO III [4], and VERITAS [5]. To explain these observations, DARM proposed that the photons are converted into axion-like-particles (ALPs) in the intergalactic magnetic fields and again into photons closer to earth. Simet, Hooper and Serpico [6] considered the scenario in which the photons convert to ALPs in the magnetic fields near the originating blazar, surviving cosmological distances as ALPs, and reconverting to photons in the magnetic field of the Milky-way galaxy by the mechanism similar to that of a Sikivie helioscope [7]. The question of interest here is: using the sun as a source, can these hypotheses be tested at the needed sensitivities? Many previous experiments were motivated by the desire to search for Peccei-Quinn (PQ) axions [8] to verify the elegant solution of the strong-CP problem. In the case of PQ axions, the coupling constant to electromagnetic fields, $g_{a\gamma\gamma} \equiv 1/M$, and the mass of the axion, $m_a$, are connected by an inverse relationship, $g_{a\gamma\gamma} = Q/m_a$, where Q is a function of parameters that depend on the specific axion model. In the case of ALPs, on the other hand, the coupling and mass can be completely independent. In a recent letter [9], we demonstrated that using the axio-electric effect for detection, and annual modulation, one might hope to make a direct observation of solar ALPs using large low-background scintillation detectors. However, this technique cannot be used to test the hypothesis in the case of a null result because it involves two coupling constants both with unknown values. In this paper we consider techniques that depend only on coupling to electromagnetic fields.

A very relevant question is: in what ranges of values of $M = (1/g_{a\gamma\gamma})GeV$, the mass scale of the coupling, and axion-mass, $m_a$, might we expect to find ALPs that would explain the cosmological photon-survival problem? There is a large uncertainty in these values because of the uncertainties in the strengths and geometries of the magnetic fields involved. However, recently several groups have given ranges of couplings and ALP masses of interest in these scenarios. Burrage, Davis and Shaw [10] suggest two interesting ranges based on the scatter in x- or $\gamma$-ray luminosity relations of various AGNs. Their analyses imply that the scatter could be caused by strong ALP-photon coupling with keV photons that occur in galaxy clusters if $M \approx 10^{11} GeV$ and $m_a << 10^{-12} eV$. In another scenario, they suggest that if the conversions occur in fields close to the AGN, possible ranges might be: $M \approx 10^{10} GeV$, and $m_a << 10^{-7} eV$. DARMA gives a range of interest of $10^{11} \leq M \leq 10^{13} GeV$ [1]. Simet et al., suggest a value of about $M \approx 5 \times 10^{11} GeV$ [6]. Among these values of M, the smallest, corresponding to the strongest coupling, is very close to the present sensitivity of the CAST experiment [11], while more realistic values of M are an order of magnitude or more larger. One obvious first step would be to try to improve the background and run the CAST experiment in the vacuum mode, but that would probably only probe the strongest coupling scenario. For comparison, we consider the possibility of using ultralow background large Xe scintillation detectors. We show that even using the most optimistic value of the coupling ($M = 10^{10} GeV$) a search using such a detector is not encouraging for testing the photon-ALP hypothesis to explain the phenomena discussed in Refs. [1,6,10].

**CALCULATION OF THE CROSS SECTION**

In this section we derive the appropriate cross section in a form convenient for making numerical computations of detection rates. A general expression for the differential cross section was given by Buchmuller and Hoogeveen [12], and is written as follows:

$$\frac{\partial \sigma}{\partial \Omega} = \frac{\hbar^2 c^2}{16\pi^2 M^2} F_a^2(2\theta) \sin^2 \theta \qquad (1)$$

In equation (1), $\hbar^2 c^2 / 16\pi^2 M^2 = 2.5 \times 10^{-50} cm^2$ for a mass scale of $M = 10^{10} GeV \Rightarrow g_{a\gamma\gamma} = 10^{-10} GeV^{-1}$.

$F_a(2\theta)$, is the form factor which can be written:

$$F_a(2\theta) = \frac{Zek^2}{(1/r_0)^2 + 2k^2(1-\cos 2\theta)} \qquad (2)$$

In equation (2), $r_0$ is the atomic screening length. With a change of variables, i.e., $r_0 k \equiv \eta$, and $\cos 2\theta \equiv x$, the form factor can be written as follows:

$$F_a(x) = \frac{4\pi Z^2 \alpha \eta^4}{\left[1 + 2\eta^2(1-x)\right]^2}. \quad (3)$$

Accordingly, the cross section can be written:

$$\sigma_a = \frac{Z^2 \alpha (\hbar c)^2}{2M^2} \int_1^{-1} \frac{\eta^4 (1-x^2) dx}{\left[1 + 2\eta^2(1-x)\right]^2}, \text{ and } Z^2 \alpha (\hbar c)^2 / 2M^2 = 4.15 \times 10^{-47} cm^2 \quad (4)$$

Integrating over $x$, the final expression for the cross section is:

$$\sigma_a = 4.07 \times 10^{-47} cm^2 \left\{ (1 + \frac{1}{2\eta^2}) \ln(4\eta^2 + 1) - 2 \right\} \quad (5)$$

The solar-axion flux was given by Raffelt [13] and for $M = 10^{10} GeV$ is written:

$$d\Phi_a / dE_a = 6 \times 10^{10} E_a^{2.481} e^{-E_a/1.205} cm^{-2} s^{-1} keV^{-1} \quad (6)$$

The total rate for interactions in Xe can be integrated numerically:

$$\int_2^{15} \sigma(E_a) \Phi(E_a) dE_a = 1.83 \times 10^{-36} s^{-1} / atom \quad (7)$$

The number of Xe atoms/ kg is $4.62 \times 10^{24}$ so that the total rate then would be $0.267 / ton / y$ for $M = 10^{10} GeV$. The rate would be lower by a factor of $1.6 \times 10^{-7}$ if the less optimistic value $5 \times 10^{11} GeV$ suggested by Simet et al. [6] was used.

**LARGE LIQUID XENON DETECTORS**

While in reference [9] we considered several scintillation materials, in this work we concentrate on xenon using the fiducial mass and low backgrounds predicted by the XMASS collaboration as an example [14,15]. The XMASS project is experimenting with an 800 kg liquid Xe-scintillation detector to develop a 20-ton spherical detector with a 10-ton fiducial volume with a predicted background rate of $10^{-7} / day / keV / kg$, surrounded by photomultiplier tubes. We will use this proposed detector design as a possible scintillation detector concept for a solar ALP search. The above background scales to 0.365 background counts per ton year in the 10-keV range of the predicted solar axion spectrum. The predicted rate and background are comparable; however, to make this technique viable, the background would have to be reduced by a factor of 10, and one would need a exposure of about $100 ton \cdot y$ to make an observation for a coupling corresponding to $M = 10^{10} GeV$. It would also have to use annual modulation to make an observation to disentangle the signal from the background, whereas a CAST-type detector collects background data most of the day when not able to directly view the solar core. The above analysis clearly demonstrates that this technique would require about 10 live years of operation to reach the sensitivity of the present CAST results with Xe detectors presently under consideration.

On the other hand, a search using the 10-ton liquid-Xe detector with volume and background predicted for the XMASS detector, but using the axio-electric effect for detection (see Table I of ref. [9]), there would be about 89 counts per day in the 1.25-ton fiducial volume if both coupling constants are $10^{-10} GeV^{-1}$, with much less than 1 count per day of background. However, as stated before, while this scenario may have interesting discovery potential [9], a null experiment would have no value in placing an upper bound on ALP coupling to photons that would test the hypotheses of interest here. Nevertheless, the data from such large Xe detectors would be interesting from the point of view of the searches discussed in Ref. [9].

**CONCLUSION**

The above calculations demonstrate that it is extremely doubtful that giant scintillation detectors for solar ALPs could in the foreseeable future serve as a test of the cosmological ALP-photon hypothesis, if that test involved only coupling to photons. The operation of CAST in the vacuum mode, even with improved background and with very long running times, may not have much hope of reaching the sensitivity of $M \approx 10^{11} GeV$. This is because the detection rate scales as $(1/M)^4$, which means that the expected counting rate would be $10^{-4}$ of that at $M = 10^{10} GeV$. Accordingly, the present version of CAST itself is probably near the limit of its technology. While the conversion probability scales as, $(BL)^2$, the squares of the magnetic field and length the helioscope [7], increasing the length of a magnet much beyond $10m$ that can track the sun, and/or increasing the magnetic field much beyond $10T$, are not very realistic. On the other hand the analyses in references [1,6] suggest that the coupling strength of ALPs and photons of interest in cosmology may correspond to a mass scale more like $M = 10^{12}$ or $10^{13} GeV$. This would very probably be beyond the capability of an improved version of the present CAST experiment. However, there is a new concept being developed by the CAST collaboration that is intended reach a sensitivity corresponding to $M \approx 10^{11} GeV$ [16].

There are, however, alternate ideas suggested by Guendelman and his colleagues [17-19] that propose to exploit the effect of photon-axion splitting in inhomogeneous magnetic fields to possibly reach sensitivities of order $M = 10^{14} GeV$. This sensitivity might be feasible because the effect depends linearly on the coupling constant, $g_{a\gamma\gamma} \equiv (1/M)$, while in experiments proposed to date, the detection sensitivity depends on the coupling constant to the second power. The solar-ALP flux also depends on the 2nd power of the coupling, which makes great improvements in sensitivities solar-ALP experiments very challenging. Although experiments exploiting the physical principles suggested in [17-19] may themselves be very challenging, they deserve serious consideration.

While a new version of CAST that would be capable of increasing the sensitivity by an order of magnitude would certainly be interesting, an experiment to provide a rigorous test of the cosmological photon-ALP oscillation hypothesis might require a sensitivity of $M \approx 10^{13} GeV$. Accordingly, new approaches will be necessary to address this challenge.